\documentclass{amsart}
\usepackage{caption}
\usepackage[numbers]{natbib}
\usepackage{a4wide,setspace}
\usepackage[pdftex]{graphicx}
\usepackage{nameref,hyperref}
\usepackage{amsmath, amssymb, psfrag, tabls}
\usepackage{lineno}
\begin{document}
\begin{spacing}{1.8}
\title{The impact of cellular characteristics on the evolution of shape homeostasis}

\author{Philip Gerlee$^{1,\ast}$, David Basanta$^2$ \& Alexander R.A.\ Anderson$^2$}
\begin{abstract}
The importance of individual cells in a developing multicellular organism is well known but precisely how the individual cellular characteristics of those cells collectively drive the emergence of robust, homeostatic structures is less well understood. For example cell communication via a diffusible factor allows for information to travel across large distances within the population, and cell polarisation makes it possible to form structures with a particular orientation, but how do these processes interact to produce a more robust and regulated structure? In this study we investigate the ability of cells with different cellular characteristics to grow and maintain homeostatic structures. We do this in the context of an individual-based model where cell behaviour is driven by an intra-cellular network that determines the cell phenotype. More precisely, we investigated evolution with 96 different permutations of our model, where cell motility, cell death, long-range growth factor (LGF), short-range growth factor (SGF) and cell polarisation were either present or absent. The results show that LGF has the largest positive impact on the fitness of the evolved solutions. SGF and polarisation also contribute, but all other capabilities essentially increase the search space, effectively making it more difficult to achieve a solution. By perturbing the evolved solutions, we found that they are highly robust to both mutations and wounding. In addition, we observed that by evolving solutions in more unstable environments they produce structures that were more robust and adaptive. In conclusion, our results suggest that robust collective behaviour is most likely to evolve when cells are endowed with long range communication, cell polarisation, and selection pressure from an unstable environment.

\end{abstract}

\maketitle

\noindent {Keywords:} shape homeostasis, evolution, development, target shape

\vspace{1cm} \noindent $^\ast$ Corresponding author: gerlee@chalmers.se

\noindent $^1$Mathematical Sciences, Chalmers University of Technology and University of Gothenburg, Chalmers tv\"argata 3, 412 58, Gothenburg
 
\noindent $^2$Integrated Mathematical Oncology, Moffitt Cancer Center,  2902 Magnolia Drive Tampa, FL 33612.

\clearpage
\section{Introduction}
Many biological systems exhibit collective dynamics in which organisms interact to form complex population-level phenomena. This behaviour is perhaps most striking among eusocial insects such as ants or wasps, but occurs in all kingdoms of life ranging from bacteria to humans \cite{sumpter2006}. The dynamics of these emergent phenomena depend on communication, decision-making and the actions of the constituent organisms. 

For example, the formation of biofilm is often achieved using quorum sensing, a mechanism by which the bacteria produce an extra-cellular diffusible substance, which binds to cell surface receptors inducing the transcription of target genes (e.g. genes that control biofilm production)\cite{parsek2005}. When cellular densities are low, the concentration of the signalling substance remains low and the target genes remain inactive, while above some threshold density the behaviour is induced. 

Another well-studied example of collective behaviour is the transition from a single cell state to a multi-cellular behaviour in the slime mould \textit{Dictyostelium discoideum} \cite{loomis2012}. This transition is caused by low levels of nutrients and is facilitated by the production and secretion of cyclic AMP (cAMP), which diffuses and induces synchronised intra-cellular oscillations. Cells start migrating towards the centre of the colony and a distinct spiral pattern is formed. The resulting structure is a fruiting body, which allows some of the cells to be transported as spores to new locations. 

The formation of an adult organism from a single fertilised cell can also be viewed as a collective phenomenon, being orchestrated by growth factors, communication across cell junctions and mechanical cues \cite{friedl2009}. The outcome in this case is orders of magnitude more complex than a biofilm or a fruiting body, but still emerges from a self-organised process were extra-cellular cues interact with the dynamics of intra-cellular signalling networks. 

Our understanding of how collective behaviour is shaped by underlying actions and interactions is however still in its infancy. Progress is hampered both by the immense complexity of biological systems, where processes on many length and time-scales interact, and our lack of a theory of collective behaviour and self-organisation. Such a theory could  inform us about the cellular phenotypes and cell-cell communication that are necessary and sufficient in order for a population of interacting organisms to attain a certain structure. For example, it is known that cell death \cite{meier2000} and cell polarisation \cite{strutt2003} play important roles in the development of many organisms, but for which morphologies are these essential processes? 

{In this study we analyse how different cellular capabilities (e.g.\ motility, polararisation) and modes of cell-cell communication influence the ability of multi-cellular aggregates to form homeostatic structures.}
The aim is not to model a specific developmental system, but to provide evidence for and against the importance of certain cellular traits for achieving structural homeostasis in different environments.  {However, some of our modelling choices (e.g. 2-dimensional domain) might restrict the generality of our results.}
In addition a formalisation of the developmental process into a computational model will hopefully make it easier to theorise and form the concepts necessary for future studies.

Our results show that both short and long range cell-cell communication and cell polarisation facilitate shape homeostasis and growth into predefined structures. On the other hand cell migration, apoptosis and short range communication have a small to negative impact on the ability to achieve these targets. The evolved solutions are robust to both external and internal perturbations, and we also show that solutions evolved in a fluctuating environment can adapt better to changing external conditions, although this was not specifically selected for. 


\subsection{Previous work}
The problem of development and shape homeostasis has been studied by a number of researchers in the artificial life community. It has generally been approached by considering the challenge of generating and developing homeostatic shapes within a computational model, often with the help of a genetic algorithm to assist the exploration of the parameter space of the model. 

The first study to consider these ideas was the seminal work of de Garis \cite{DeGaris1992}, who evolved cellular automaton (CA) rules in order to generate simple predefined shapes (square, triangle). The update rules of the CA were time-dependent so that the time step influenced the dynamics, or in other words each time step had its own lookup table. {The model could evolve simple convex shapes, but in order to form non-convex shapes (e.g.\ an L-shape), a morphogen was introduced. A morphogen is a chemical substance that regulates development, and in that model it was assumed that it is passed on in decreasing amount from parent to daughter cell}. The cells were endowed with the capability to sense the morphogen gradient, and this input signal could switch the cells between different behaviours. {The evolution of pattern generation in CAs has been improved in recent years by employing an instruction-based approach, see e.g. the work by Bidlo et. al \cite{Bidlo2012, Bidlo2016}}.

Amongst the work that built upon the ideas of de Garis was the one performed by Basanta and colleagues \cite{Basanta2008}. They extended the model of de Garis in a number of directions: firstly they considered a three dimensional system, secondly they did not define specific shapes, but rather looked for patterns that retained their shape after development had occurred, and lastly they allowed for both more inputs to the cell and ranges of cell behaviours. The cell behaviour was a function of the number of neighbouring cells, the time elapsed and additionally the number of cell divisions the cell had gone through, whereas the range of cell action was extended to account for movement and apoptosis (death). They showed that a variety of different strategies for shape homeostasis evolved, and that they differed in their ability to self-repair upon wounding, the most efficient being those structures that had a directional flux of cells, much like real tissues. 

In an effort to more closely resemble biological development a number of researchers have opted for dynamical networks to determine cell actions rather than the rule-based dynamics of de Garis and Basanta et al. Eggenberger \cite{Eggenberger1997} used a gene regulatory network within each cell in order to evolve stable shapes triggered by morphogenetic signals in the environment. The protein expression of one cell could influence the genetic network of neighbouring cells, which effectively allowed for cell-cell communication. 

Streichert et al. \cite{Streichert2003} used an off-lattice model, in which each cell was equipped with a simple gene network in the form of a Boolean network. The role of these boolean networks was to map external stimuli to the behaviour of the cell (e.g. cell division or death). They showed that a simple network consisting of only two genes was sufficient for homeostasis, but that the addition of a death-signal was necessary to achieve self-repair. They also compared the performance of boolean networks with a continuous non-linear model of gene-regulation and found that the former achieved better results in terms of fitness. 

The inherent complexity involved in evolving complex patterns was tackled by Roggen and Federici \cite{Roggen2004} by subdividing the developmental process into subroutines that were governed by distinct genetic networks. The intra-cellular dynamics were modelled using feed-forward neural networks with an extensive list of inputs (e.g. current cell state, external growth factor, state of neighbouring cells) and the output being the cell type, production of growth factor, and direction of cell division. The evolutionary algorithm proceeded by first selecting a network for a prepattern, which was then fixed, and then a new network was evolved for the second embryonic subroutine. This method increased the efficiency of the process, but also makes it more difficult to find parallels with biological reality, since each successive subroutine is fixed and immutable. 

The evolution of specific target shapes such as hollow spheres or solid cubes was investigated by Andersen et al. \cite{Andersen2009} in the context of a model with a continuous gene regulatory network. The cells could sense the protein expression of neighbouring cells and the concentration of an external growth factor that exhibited a gradient in the computational domain. They carried out a detailed analysis of the topology of the evolved intra-cellular networks, and showed that  the same end result could be achieved with different genotypes, that followed distinct developmental trajectories. They also demonstrated that the ability to heal wounds emerged even though it was not part of the fitness function. 

In a more recent study by Chavoya et al. \cite{Chavoya2010} genetic algorithms were used in order to reproduce the pattern of the French flag \cite{Wolpert1969}. In their model the dynamics in each cell was controlled by an artificial gene regulatory network \cite{Banzhaf2003}, where each structural gene corresponded to a CA lookup table. In each time step and for each individual cell the lookup table of the gene with the highest activation was used for updating the model. The networks were driven by morphogenetic gradients in the environment, no cell communication was included, and the cell action was restricted to division. 

{The French flag problem was also investigated by Miller \& Banzhaf using developmental cartesian genetic programming \cite{Miller2003}. In this framework each cell encodes a directed and acyclic graph that maps cellular input (chemical concentrations, state of neighbouring cells) to an action such as cell division or a change in cell state. The mapping was represented as a string of integers and in order to achieve a genotype that would grow into the French flag pattern a genetic algorithm was applied. Using this framework the French flag could be reproduced as well as more intricate patterns.}

The influence of noise on development was investigated by Joachimzak \& Wrobel \cite{Joachimczak2012} in the context of the French flag problem. They used a 3-dimensional off-lattice model in which the cell behaviour is encoded by a genetic regulatory network and cells interact physically by pushing and adhering. In that setting they could show that networks that had evolved under noise in the synthesis rate of proteins were more, as expected, more robust to this type of intra-cellular noise, but were also more robust to extra-cellular noise in terms of wounding. Notably, networks that had evolved in noisy conditions under-performed when the noise was removed, suggesting that stochasticity had become a necessary component for their normal functioning. 

The impact of genetic operators such as gene duplication and transposition on the evolution of shape generation was investigated by Schramm et al. \cite{Schramm2012}. In that model the transcription factors produced by the regulatory network were allowed to diffuse into the computational domain, potentially influencing not only neighbouring cells, but also distant cells. Cell action was limited to division and death, and the goal was to evolve an elongated multi-cellular structure, which turned out to be facilitated by redundancy in the genome.

Gerlee et al. \cite{Gerlee2011} investigated the evolution of cell mono-layers and found that the choice of fitness affected the outcome. They used two different schemes, one which used a constant fitness evaluation where each individual was tested for 200 time steps (constant), and another which increased the evaluation time for each successive generation (from 50 to 200, termed incremental). Analysis of the solutions provided by an evolutionary algorithm (EA) showed that the two evaluation methods give rise to different types of solutions to the problem of homeostasis. The constant method leads to near optimal solutions, which rely on a very high rate of cell turn-over, and this is achieved by fine-tuned balance between cell birth and death. The solutions from the incremental scheme on the other hand behave in a more conservative manner, only dividing when necessary, and generally have a lower fitness. In order to test the robustness of the solution they were subjected to environmental stress, by wounding the tissue, and to genetic stress, by introducing mutations. The cell types with high turn-over were more robust with respect to wounding, healing faster and more accurately. 


\subsection{Summary of previous work}
In summary, previous attempts at mathematically modelling shape homeostasis have been varied both in scope and method. Some studies have focused on the dynamics of intra-cellular signalling, while others have centred on the emergence of wound healing and mutational robustness. The model architecture has also differed, ranging from CA to network based approaches with a range of cell actions (proliferation, apoptosis and migration) and means of cell-cell communication. {Despite many previous studies in this field, to our knowledge, no study has looked at both shape generation and arbitrary homeostatic structures within the same computational model.} The different approaches are summarised in table \ref{tab:sum}, which highlights the diversity of the field.


\begin{center}
\begin{table}
    \begin{tabular}{ | l | p{1.8cm} | p{1.5cm} |p{2cm} |p{0.95cm} | p{1.3cm} | p{1.7cm} |}
    \hline
    Study & Shape/ homeostatic & Rule /Network-based &  Long/short-range communication & Cell action & Division polarity & External morphogen\\ \hline \hline
    de Garis \cite{DeGaris1992} & Shape & Rule & Both & P & Yes & No  \\ \hline
    Basanta \cite{Basanta2008} & Homeostatic & Rule &  Short & PDM & Yes & No \\ \hline
    Eggenberger \cite{Eggenberger1997} & Homeostatic & Network  & Short & PD & No & Yes \\ \hline
    Streichert \cite{Streichert2003} & Homeostatic & Network  & Short & PD & No & No \\ \hline
    Federici \cite{Roggen2004} & Shape & Network  & Both & P & Yes & No \\ \hline
    Andersen \cite{Andersen2009} & Shape & Network  & Short & PD & No & Yes \\ \hline
    Chavoya \cite{Chavoya2010} & Shape & Both &  Neither & P & Yes & Yes \\ \hline
   Miller \cite{Miller2003} & Shape & Network &  Both & PD & Yes & No \\ \hline
    Joachimczak \cite{Joachimczak2012} & Shape & Network  & Both & PD & Yes & Yes \\ \hline
    Schramm \cite{Schramm2012} & Shape & Network  & Short & PD & Yes & Yes \\ \hline
   Gerlee \cite{Gerlee2011} & Shape & Network  & Short & PDM & No & Yes \\ \hline
        \end{tabular} \caption{A summary of previous attempts at modelling shape development and homeostasis. Long range communication refers to a diffusible substance that the cells can emit and sense, while short range refers to the cells having information about their local neighbourhood. The cell actions refer to proliferation (P), death (D) and migration (M).} \label{tab:sum}
        \end{table}
\end{center}

\subsection{Aim of this study}
The aim of the current study is to make sense of this veritable jungle of results and  quantify the impact of modelling choices on the shapes that can be evolved. We will focus on three key aspects of the models: cell-cell communication, cell actions and division polarity. {To be clear we take these to mean the following: \textit{cell-cell communication} is the ability of cells to communicate by secreting and sensing a chemical substance, \textit{cell actions} are the range of possible behaviours endowed to the cells (e.g. migration, cell death) and \textit{division polarity} is the ability of cells that divide to place the offspring in a desired neighbouring location (and not at a random neighbouring site).} 
By varying these modelling choices we explore how they influence the ability to achieve two different evolutionary objectives: developing into a predefined shape and growing and maintaining an arbitrary shape (homeostasis). {Naturally other modelling choices such as the choice of cellular decision making (network vs.\ rule-based) and the type of computational domain (2D vs.\ 3D and discretised lattice vs.\ off-lattice models) affect the results. However, not every aspect of modelling can be analysed within a single study and we therefore make the following concessions to generality: (i) we will consider a network model of cell dynamics (and not a rule-based CA), (ii) make use of the discretised space  where cells reside on a two-dimensional square lattice, and (iii) we will also disregard the presence of an external morphogen. 
The motivation for these choices is partly biological and partly computational. A two-dimensional lattice was chosen since it is considerably faster to solve the diffusion equation on such a domain, whereas intra-cellular networks were chosen (instead of a rule-based model) since they closer resemble the dynamics in real developmental systems \cite{Levine2005}. The current setup is of course far from a biological system, but if we in future studies hope to generalise the results to biological systems then a model which resembles the biological system if preferable. Lastly, we disregard external morphogens since {many} real developmental systems are not endowed with external morphogens, but are fully self-organised and have to generate their own morphogenetic gradients.}

We will also take this study one step further and consider the effect of an unstable environment on the evolution of homeostasis, which to our knowledge has not been previously considered. This is achieved by the introduction of harmful particles that are released from the boundary of the computational domain. These particles move according to a random walk and when they come into contact with a cell it dies and the particle is annihilated in the process. 


\section{Methods} \label{sec:methods}
We have investigated the emergence and robustness of a developmental process within a multi-cellular context and focused on the impact of a number cellular characteristics (or equivalently modelling choices) on these processes. We  considered two distinct targets for the fitness function of the evolutionary algorithm (EA): a predefined non-convex shape and maintaining any non-trivial shape over time {(defined in section 2.7.1)}. 

{For each of these targets we also varied the model properties. With regards to cell-cell communication we considered the presence/absence of two types of signalling that is mediated by chemical substances that we term growth factors (GFs) \cite{Lovicu2011}: short range GF (SGF) and long range GF (LGF). Both of these chemicals are produced and can be sensed by the cells. The short range SGF is deposited locally and subject to decay, and the long range LGF is deposited locally and decays, but is in addition subject to diffusion (see section 2.4 for details). We will consider 3 variations in growth factor inclusion: only SGF, only LGF and  SGF and LGF acting together.}


{With regards to cell action we included three basic cellular mechanism that are thought to be important in development \cite{Kurosaka2008}: cell proliferation (P), death (D) and migration (M). We will investigate the biologically relevant permutations of these mechanisms: P, P+D, P+M and P+D+M.}

{Cell polarity, i.e.\ the ability of cells to behave in a spatially anisotropic fashion, is known to be an important mechanism in development \cite{Cove2000}, yet it has not been considered in all previous models (see Table 1). We therefore consider the presence/absence of cell polarisation as a variable property and run the model both with and without this feature (denoted POL and $\neg$POL).}

{Lastly, we will try all possible combinations the above model choices in two types of environments, one where cells remain  unperturbed and an alternative one where harmful particles are introduced at a fixed rate (see section 2.5 for details).}

This means that in total we will investigate the dynamics of the model for 2 (targets) $\times$ 3 (GF settings) $\times$ 4 (cell actions settings) $\times$ 2 (polarisation settings) $\times$ 2 (environmental settings) = 96 distinct configurations with the aim of determining the impact of cellular properties on achieving the different targets. For brevity we will refer to a given cellular setting using a binary string $S=(s_1s_2s_3s_4s_5)$, where $s_i \in \{0,1\}$. Here $s_1$ denotes the presence/absence of short range growth factor, $s_2$ corresponds to long range growth factor, $s_3$ corresponds to cell death, $s_4$ to motility and lastly $s_5$ denotes the presence/absence of cell polarisation. For example the string $S=(01010)$ denotes a setting in which the cells are endowed with LGF and can move.


\subsection{Modelling framework}
We considered a  population of cells residing on a two-dimensional discrete lattice of $50 \times 50$ grid points. 
Each grid point can be either empty or occupied by a cell. {The boundaries of the domain are fixed and cells cannot leave the lattice}.
Each of these cells on the lattice contains a signalling network, modelled using a recurrent neural network, that dictates its behaviour. In the most extended version of the model the cells are able to sense the local concentration of SGF and LGF. {The cells are uniform in the sense that they all contain the same signalling network (genotype), but two cells might react differently to identical stimuli since their networks might be in different states.}
The GFs are not externally supplied but produced by the cells themselves in an amount that is controlled by the signalling network. 
The network also determines the behaviour of the cell, if it proliferates, moves or dies, as well as controling the polarisation -- which influences the direction of migration and cell division. 

\subsection{Signalling network}\label{sec:network}
\begin{figure}[!htb]
   \centerline{\includegraphics[width=12cm]{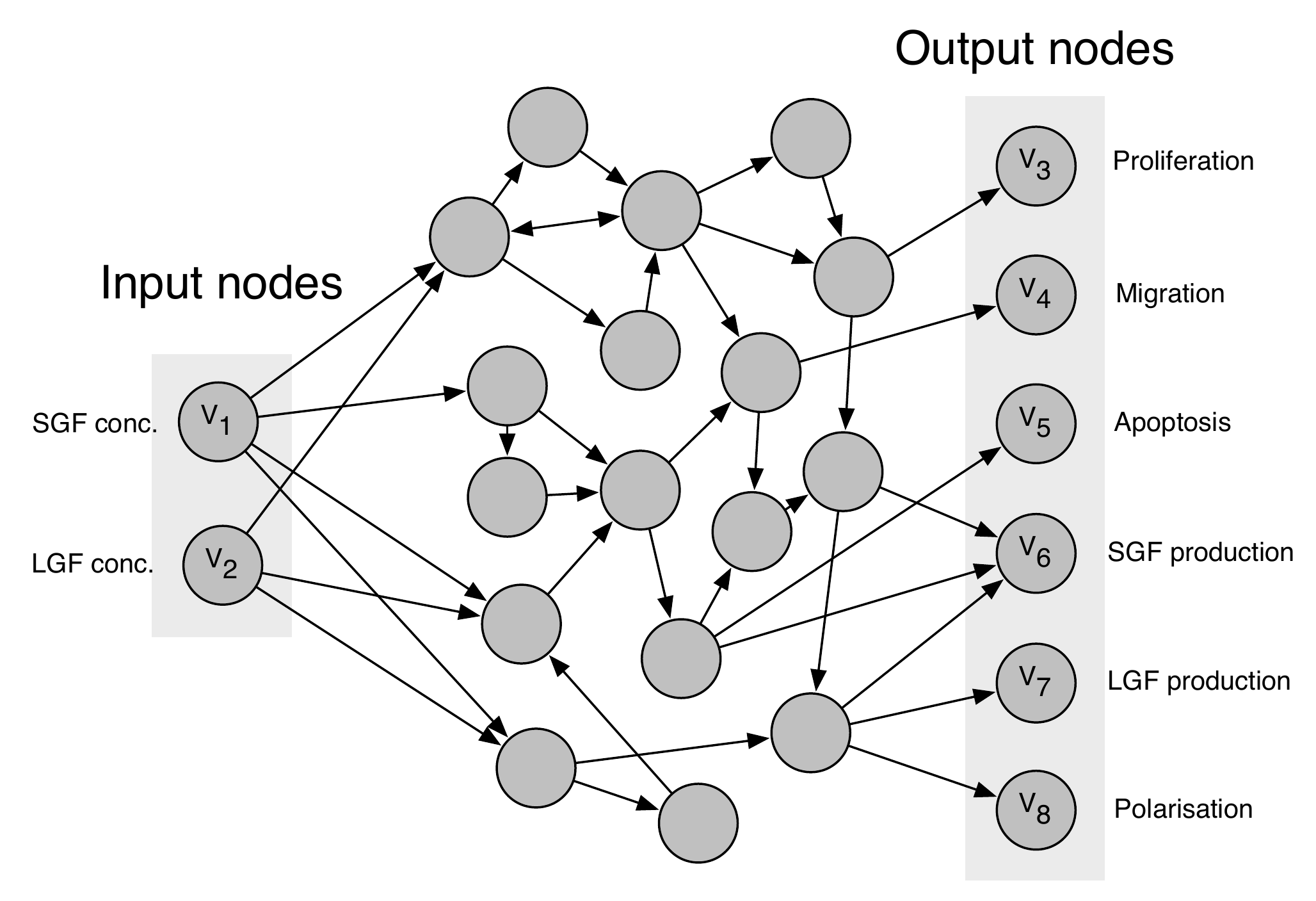}}
\caption{\label{fig:net}{An example of a signalling network which determines the behaviour of the cells.  The values of the input nodes are determined by the {GF concentrations at the location of the cell}, while the state of all other nodes is determined dynamically by eqn.\ \eqref{eq:update}. The output nodes determines the cell action (proliferation, migration, apoptosis), the growth factor (GF) production and direction of polarisation. For clarity only a subset of the links are shown.}}
\end{figure}

The network inside each cell consists of $n=25$ nodes (see fig. \ref{fig:net}), and the state of each node is updated each time step according to
\begin{equation}\label{eq:update}
V_{i}(t+1) =  g\left(\sum_{j=1}^{n} w_{ij}  V_{j}(t) \right) 
\end{equation}
where the matrix element $w_{ij} \in [-1,1]$ describes the connectivity between node $i$ and $j$, and 
\[g(x) = \frac{1}{1+e^{- \beta x}} \]
is a transfer function that ensures that $V_i(t) \in [0,1]$. {The number of nodes was chosen in such a way that the range of possible dynamical behaviours would be large enough to allow for complex dynamics, and at the same small enough so that the parameter space searched by the evolutionary algorithm would not be too large. The choice of $n=25$ reflects this trade-off.} {It would be possible to use network with variable size \cite{Trefzer2013,Nichele2014}, but for simplicity we use a fixed size.}

\subsection{Cell dynamics}
A subset of the nodes in the network are assigned as input nodes. They lack incoming connections and their values are instead defined by extra-cellular conditions. The network contains two such nodes: $V_1$ senses the SGF, $V_1(t) = S_{i,j}(t)$, and $V_2$ senses LGF, $V_2(t) = L_{i,j}(t)$, where $S_{i,j}(t)$ and $L_{i,j}(t)$ denote the time and space dependent extra-cellular concentrations of SGF and LGF respectively.  


Another subset of the nodes in the network determines the behaviour of the cell and are called output nodes. We consider six such nodes, out of which three control cell action, two control GF production and the last one cell polarisation. 

The three cell action nodes are each associated with a specific behaviour: $V_3$ -- proliferation, $V_4$ -- migration and $V_5$ -- apoptosis. At each time step the cell carries out the action whose node value is highest, unless $V_{3,4,5} < 1/2$, in which case the cell remains quiescent. If proliferation is chosen an internal counter is incremented and when it has reached a certain value $t_p = 10$, corresponding to the time of the cell cycle, the cell divides (and the counter is reset). The daughter cell is placed in a neighbouring grid point {(using a von Neumann neighbourhood)} dictated by the polarisation node (see below) and receives a copy of the parent network. If polarisation is disabled the offspring is placed at random among the neighbouring grid points and {if that grid point is occupied cell division fails. If migration is chosen the cell moves in the direction decided by the polarisation node. If polarisation is disabled movement is random, and fails if the chosen grid point is occupied.} Lastly, if apoptosis is chosen the cell dies and leaves an empty grid point behind. {The instant nature of cell death in the model (as compared to cell division) is motivated by biological considerations: apoptosis can occur on the time scale of minutes, while cell division is a process on the scale of hours \cite{meier2000}.} {For more details on the cell dynamics we refer to \cite{Gerlee2011}.}

Node $V_6$ controls the rate of SGF production $\sigma_{i,j}(t)$, while $V_7$ controls the rate of LGF production $\lambda_{i,j}(t)$.

The polarisation node influences both the proliferation and migration dynamics by dictating the direction of cell division/movement. {If $V_8 \leq 0.25$ the preferred direction is north on the lattice, if $0.25 < V_8 \leq 0.5$ it is south, if $0.5 < V_8 \leq 0.75$ it is east and otherwise west. However, if the preferred direction is occupied a new random direction is chosen. The same applies if the cell resides next to the boundary and polarisation dictates the cell to move/place offspring outside the domain.}

\subsection{Growth factor dynamics}
The SGF is released by the cell at location $(i,j)$ at rate $\sigma_{i,j}(t)$ and decays at a constant rate $\delta_s = 1/2$ {(corresponding to a half-life of 1.4 cell cycles)}. The concentration $S_{i,j}(t)$ at each grid point $(i,j)$ therefore obeys the ordinary differential equation
\begin{equation}\label{eq:SGF}
\frac{dS_{i,j}(t)}{dt} = \sigma_{i,j}(t) - \delta_s S_{i,j}(t), \ i,j = 1,2,...N. 
\end{equation}
The LGF behaves in a similar way, but is also subject to diffusion with diffusion constant $D=1$ (in dimensionless units) and decay rate $\delta_l = 0.1$. The LGF concentration therefore obeys the equation
\begin{equation}\label{eq:LGF}
\frac{\partial L_{i,j}(t)}{\partial t} = D\nabla^2 L_{i,j}(t) + \lambda_{i,j}(t) - \delta_l S_{i,j}(t), \ i,j = 1,2,...N. 
\end{equation}
where the Laplacian is approximated using a five-point stencil, and we impose no-flux boundary conditions. {The parameters for the LGF were chosen so that a single cell producing LGF (at the centre of the domain) could give rise to a gradient in LGF across the domain.} 

\subsection{Death particle dynamics} 
In the simulations where we studied the role of an unstable environment, the harmful death particles are released from the boundary of the computational domain. Each time step, $n_p$ new particles are released from random points on the boundary, and particles already in the system move one grid point in a random direction. The particles are assumed not to interact with one another, i.e.\ no collisions occur. If a particle arrives at a grid point inhabited by a cell, the cell dies and the particle is annihilated in the process.

\subsection{System dynamics}
The initial conditions of the system is a zero GF concentrations in the entire domain, and a single cell with a given signalling network  at the centre of the grid at position $ (i,j) = (N/2,N/2)$. {All the nodes of the the cell's network are initialised with $V_i=0$ for all $i$}.
At the start of each time step the SGF and LGF concentrations are updated by numerically solving  equation \eqref{eq:SGF} and \eqref{eq:LGF}. If the simulation considers an unstable environment then new death particles are introduced into the system and the positions of the existing particles are updated. All the cells on the grid are then updated {asynchronously (in a random order)} as follows:
\begin{enumerate}
\item If a death particle resides in the same grid point as the cell, the cell dies and the particle is annihilated
\item The {GF concentrations at the location of the cell} $(S_{i,j}(t),L_{i,j}(t))$ are sampled and the signalling network is updated according to equation \eqref{eq:update}
\item The cell produces SGF and LGF according to the output of the network
\item The actions associated with the chosen phenotype are carried out and the grid is updated accordingly
\end{enumerate}

\subsection{Evolutionary algorithm}

In order to to find a signalling network (i.e.\ a connection matrix $w$) that gives rise to a multi-cellular structure exhibiting shape homeostasis we make use of an evolutionary algorithm (EA)  \cite{banzhaf1998}. EA are versatile evolution-inspired optimization algorithms. The design of an EA has been abundantly described in literature \cite{mitchell1998}, but in general the two key aspects of a EA are the description of the solution to the problem that needs to be optimized and the fitness function that determines whether a solution is a good one. In our case the EA will evolve a population of solutions in which each solution describes a connectivity matrix $w_{ij}$. The fitness function will discriminate between these solutions by assigning a high fitness to those, who when seeded into a single cell, and subject to the system dynamics described above, will give rise to a multi-cellular structure that achieves a predefined target shape (homeostasis or a particular shape). Below we detail the different fitness functions.

\subsubsection{Shape homeostasis}
The aim of the EA in this case is not to recreate a specific shape, but any shape that is stably maintained. With an eye on computational efficiency, we defined shape homeostasis so that it could be assessed within reasonable time. We made the following choice: the shape of the multi-cellular structure should ideally be identical after $T/2$ and $T$ time steps, where we have chosen $T=200$ (measured in cell cycles). The fitness function of an individual in the EA (corresponding to a matrix $w$) is therefore defined as

\begin{equation}\label{eq:fitness}
F_1(w) = 1 - \frac{1}{N^2}\sum_{i,j}^N \delta(C_{i,j}(T/2),C_{i,j}(T))
\end{equation}
where $C_{i,j}(t) = 1$ if a cell is present at $(i,j)$ at time $t$ and zero otherwise, {and the function $\delta(i,j)$ is defined according to}
\[\delta(i,j) = \left\{
  \begin{array}{lr}
    0 & : i=j\\
    1 & : i \neq j
  \end{array}
\right. 
\]
{In order to avoid trivial solutions (those that correspond to an empty or completely occupied grid) we set the fitness of networks that generate either an empty or completely full grid equal to $F_1=0$.}

\subsubsection{Predefined shape}
In line with the work of de Garis we chose the predefined shape to have an L-shape (see fig.\ \ref{fig:unpert}A), which  at least in the framework of de Garis was the most difficult to achieve. We encoded the desired shape in a function $P(x,y)$ such that 
\[P(i,j) = \left\{
  \begin{array}{ll}
    1 & : {\rm if\ } P(i,j)\ {\rm is\ part\ of\ the\ target\ shape}\\
    0 & : {\rm otherwise}
  \end{array}
\right.  
\]
Again we make use of the function $\delta(i,j)$ and write the fitness as
\begin{equation}\label{eq:fitness}
F_2(w) = 1 - \frac{1}{N^2}\sum_{i,j}^N \delta(P(i,j),C_{i,j}(T))
\end{equation}
Also in this case we circumvent solutions corresponding to an empty grid by letting $F_2 = 0$ if no cells are present. {One could consider more flexible fitness functions that allow for rigid transformations and rotation of the target pattern, but here we focus on the simple case of a fixed shape.}
{Lastly, we note that for both fitness functions we thus have that the minimal value is 0 and the maximal value is 1.}

\subsubsection{Selection scheme}
\begin{figure}[!htb]
   \centerline{\includegraphics[width=10cm]{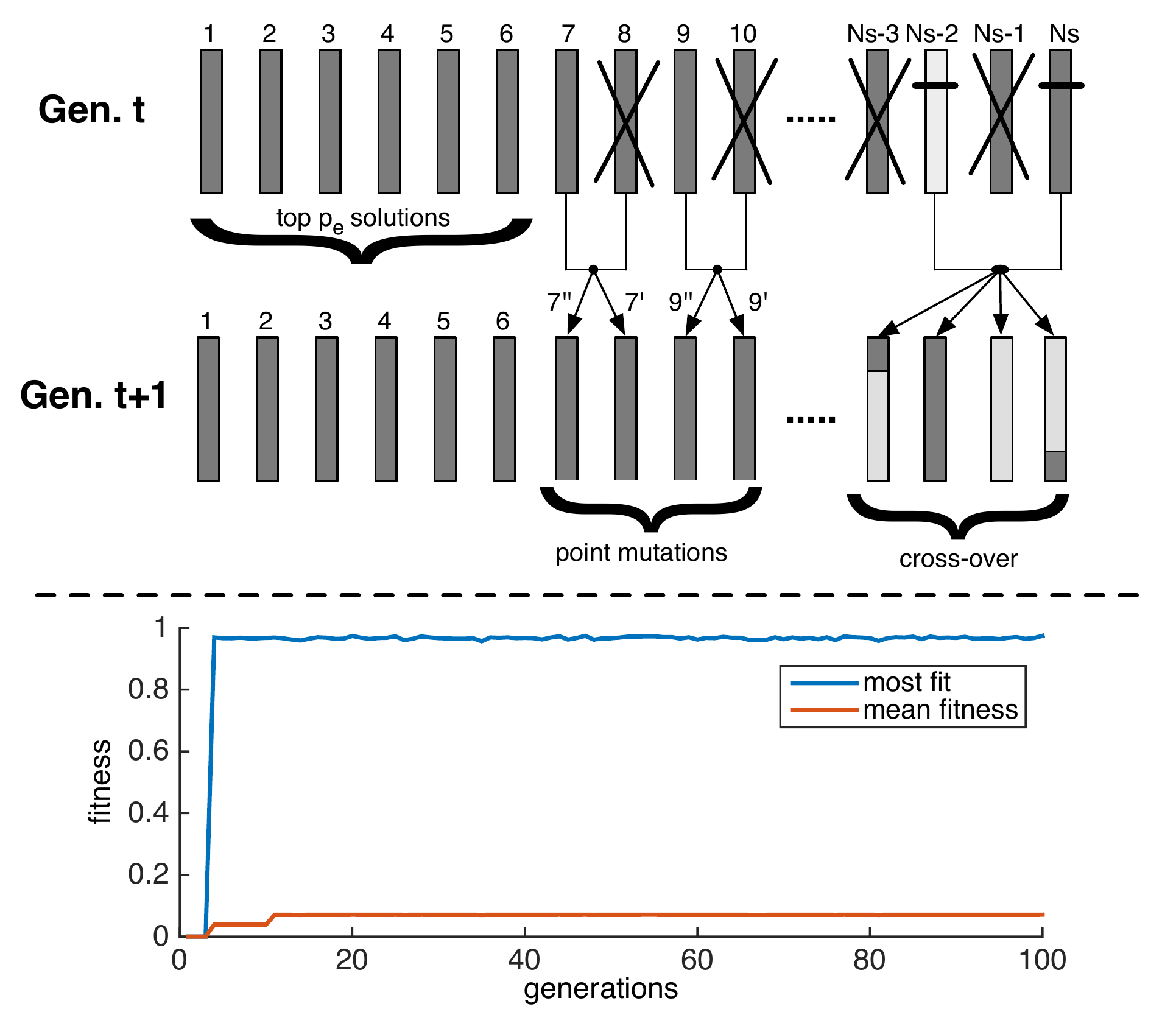}}
\caption{\label{fig:EAscheme}Top panel shows a schematic of the evolutionary algorithm. In each generation the top $p_e$ most fit solution are carried unaltered into the next generation. The remaining genotypes are paired off in tournaments where the winners give rise to offspring either by point mutations or cross-over. Bottom panel shows the fitness of the most fit individual and the average fitness in five independent runs of the evolutionary algorithm (setting 10000 in an unstable environment). The reason for the fluctuations is that fitness evaluation is stochastic and hence the fitness of a solution might vary from generation to generation. }
\end{figure}

The EA consists of a population of candidate solutions, which are subject to a selection process driven by the previously described fitness function. {We initialise the EA with a population of random individuals, given by matrices $w$ with random entries taken from a uniform distribution in the interval $[-1,1]$} (see section \ref{sec:network}). We have chosen a tournament based selection process, which also makes use of a low degree of elitism \cite{banzhaf1998}. {This means that the top $p_e = 5 \%$ fraction of the solutions are carried unaltered into the next generation. This scheme guarantees that the maximum fitness always increases, but can be detrimental if $p_e$ is chosen too large.} The rest of the population engages in tournaments.  Four solutions are picked at random and are compared in pairs, and this generates two winners and two losers. The winners are carried over to the next generation, while the losers are replaced by the offspring of the winners. The offspring is either generated by single point mutations to the parents genotypes (occurs with probability 1/2) or by crossing-over their two genotypes (with complementary probability 1/2). This process is repeated until all solutions in the population have engaged in a tournament, and this constitutes one generation in the evolutionary algorithm. See fig.\ \ref{fig:EAscheme} for a graphical representation of the selection process and the change in fitness during a run of the EA (for setting 1000 in an unstable environment). Both the fitness of the most fit solution and the average fitness increases over time, which shows that the ability to achieve homeostasis increases during the selection process.

{The point mutations are implemented by picking one of the matrix entries $w_{i,j}$ at random (with uniform probability $1/n^2$) and adding to it a random number chosen from a normal distribution with $\mu=0$ and $\sigma^2=0.7$.} The cross-over is implemented by mapping the matrix entries of both parents into two vectors, picking one random index and swapping the contents above and below this point. In most cases this is disruptive, but in some cases it generates a novel solution with a higher fitness than the parents. {More sophisticated methods for evolving network are available (see e.g.\ \cite{Stanley2002}), but we opt here for a simpler approach.}


\section{Results}
In order to characterise the impact of different modelling choices on the evolution and properties of morphogenesis we ran the EA 50 times for each setting of the model. In total $50 \times 96 = 4800$ runs each of which lasted 20 generations and contained a population of 25 candidate solutions. From each EA-run we collected the fittest solution for further analysis. {The population size of each EA and the number of generations is small, but was chosen this way in order to explore the large number of different model settings using limited computational resources.} {Longer runs with the EA only increase the mean fitness marginally.} This fact is illustrated in fig.\ \ref{fig:EAscheme}, which shows the most fit and the mean fitness of the population as a function of the number of generations. After an initial increase the fitness remains almost stationary and fluctuates at a value close to $F=1$. Fluctuations are due to the fact that fitness evaluation has a stochastic component (e.g.\ placement of daughter cells is random).

{In order to ensure that evolution is in fact taking place in this system we compare the fitness of the initial randomised solutions with those obtained after 20 generations of evolution. Averaging across all settings we find that in the case of shape homeostasis the mean fitness of the initial solutions was $\bar{F}=0.05$  compared to 0.53 for the evolved ones. Comparing the number of viable solutions (with $F>0$) we find that 5 \% of the initial solutions have non-zero fitness compared to 63 \% among the evolved ones. The corresponding numbers for the target shape are $\bar{F}=0.04$ vs.\ 0.63 and 5 \% vs.\ 93\%. This clearly shows that evolution is taking place for both fitness functions.}

In the following subsections, we first discuss the results obtained in an unperturbed environment (no death particles), and then move on to the perturbed case. In both sections we will start with a qualitative analysis of the different solutions and then move onto a statistical analysis in which we try to understand the impact of the different modelling choices. Lastly we will analyse the effect of mutations and wounding, and how this relates to the modelling choices. 

\subsection{Unperturbed case}

\subsubsection{Qualitative analysis}
\begin{figure}[!htbp]
   \centerline{\includegraphics[width=10cm]{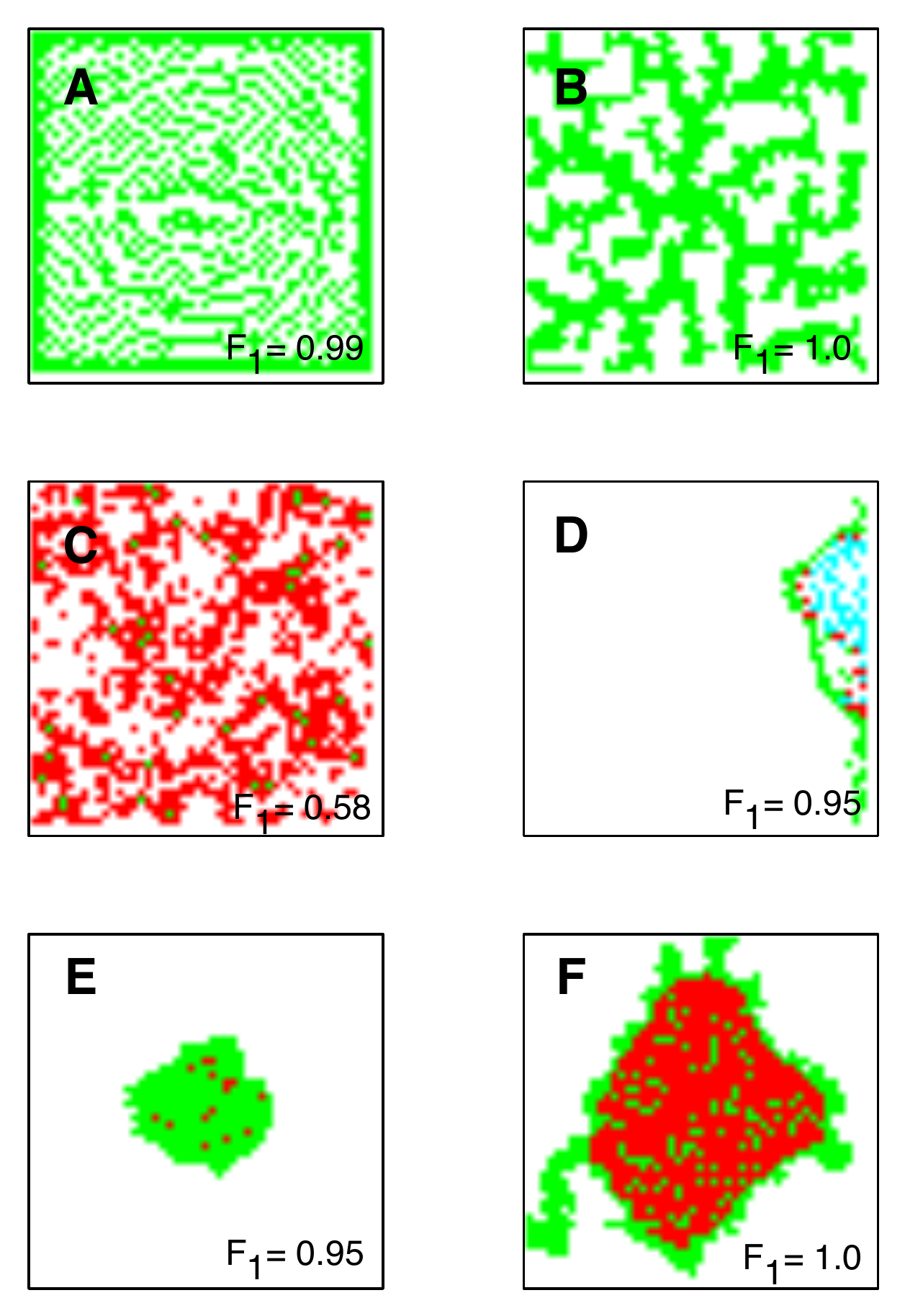}}
\caption{\label{fig:unperh}Examples of shape homeostasis in an unperturbed environment and their corresponding fitness $F_1$. A and B are solutions which are static in the sense that after they are formed no more cell divisions occur. C and D on the other hand are examples of dynamic solutions where cell division is balanced by cell death. Lastly E and F are examples of faux homeostasis where the structure grows very slowly. Green cells are quiescent, red are in the proliferative state and cyan cells are moving. {All simulations are shown after 200 cell cycles have elapsed}. The binary numbers in the top right corners show the model setting (see section \ref{sec:methods}). Movies of all panels can be found at: \href{http://www.youtube.com/playlist?list=PLfj1iGvMEpeV3J5fOXbntTxDWwlzRHUSp}{http://www.youtube.com/playlist?list=PLfj1iGvMEpeV3J5fOXbntTxDWwlzRHUSp}}
\end{figure}

\textbf{Shape homeostasis.}
We start by looking at a few examples from when the target is shape homeostasis (see fig.\ \ref{fig:unperh}). Although not evident from a single time point of the system, we can discern two broad classes of solution: static and dynamic solutions. The static solutions (fig.\ \ref{fig:unperh}a and b) grow into a fixed multicellular structure which then becomes quiescent in order to maintain its shape, whereas the dynamic solutions (e.g. fig.\ \ref{fig:unperh}c-f) form a structure which is continuously modified, whilst still (approximately) maintaining its shape. Note that for this to be possible the cells need to be endowed with cell death, and hence this behaviour is only found in those simulations. The two static solutions shown in fig.\ \ref{fig:unperh} limit their growth using LGF, which in panel b leads to a striking branched morphology. {This is similar to what is seen in diffusion limited growth \cite{Ben1998}, in which a substance diffusing from outside the growing colony limits growth. In this case, the LGF is being produced by the colony itself and acts in an opposite manner -- it inhibits growth. The net effect is however the same. Cells that are located on the tip of branches sense lower levels of LGF (compared to cells close to the centre) and consequently divide at a higher rate. This positive feedback leads to instabilities and eventually to branching behaviour.} {Note that both solution a and b make use of the fact that the domain is bounded which allows of a build up of LGF. In an unbounded domain these solutions would most likely continue growing}.

Among the dynamic solutions there is a subclass of solutions which never leave the developmental stage and continue to grow slowly, but indefinitely. This is a type of faux homeostasis, which appears as a solution because of our limited evaluation time, but interestingly such a solution exists in nature in e.g.\ the human colon, which grows at diminishing rates throughout life. 

Two examples of this slow-growing class are shown in fig.\ \ref{fig:unperh}e and f, that employ different strategies. The solution in E contains mostly quiescent cells (green) that occasionally become proliferative (red), but only if the cell sits at the perimeter will it actually divide. The other solution (panel F) contains mostly proliferating cells, but they are enclosed by a rim of quiescent cells. 

\begin{figure}[!htbp]
   \centerline{\includegraphics[width=10cm]{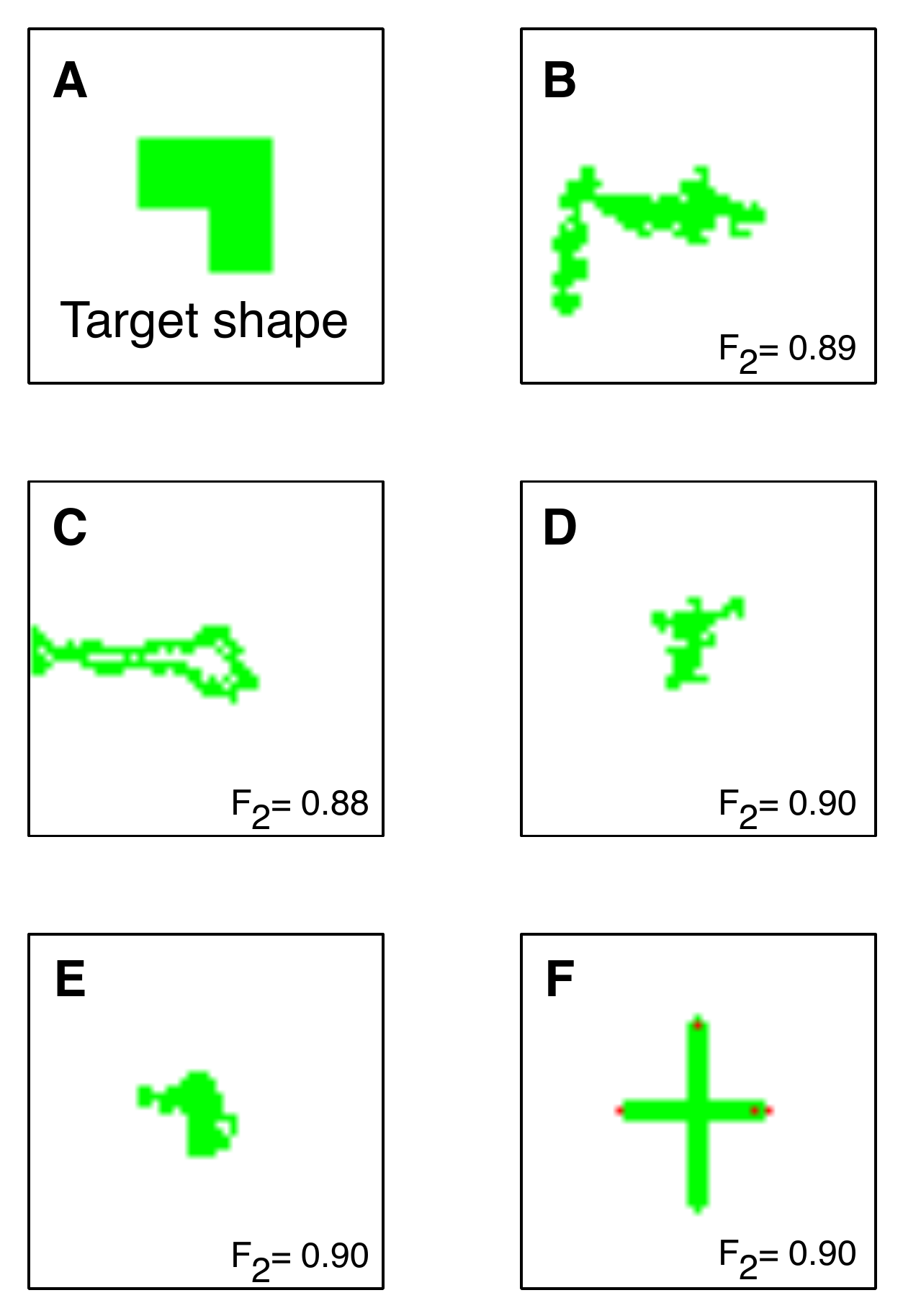}}
\caption{\label{fig:unpert}The target shape (A) and some example of target shape homeostasis in an unperturbed environment and their corresponding fitness $F_2$ (B-F). Green cells are quiescent and red are proliferating.  {All simulations are shown after 200 cell cycles have elapsed}. The binary numbers in the top right corners show the model setting (see section \ref{sec:methods}). Movies of all panels can be found at: \href{http://www.youtube.com/playlist?list=PLfj1iGvMEpeUH8R0nP0vl4hnDHBlLciYD}{http://www.youtube.com/playlist?list=PLfj1iGvMEpeUH8R0nP0vl4hnDHBlLciYD}}
\end{figure}

\begin{figure}[!htbp]
   \centerline{\includegraphics[width=10cm]{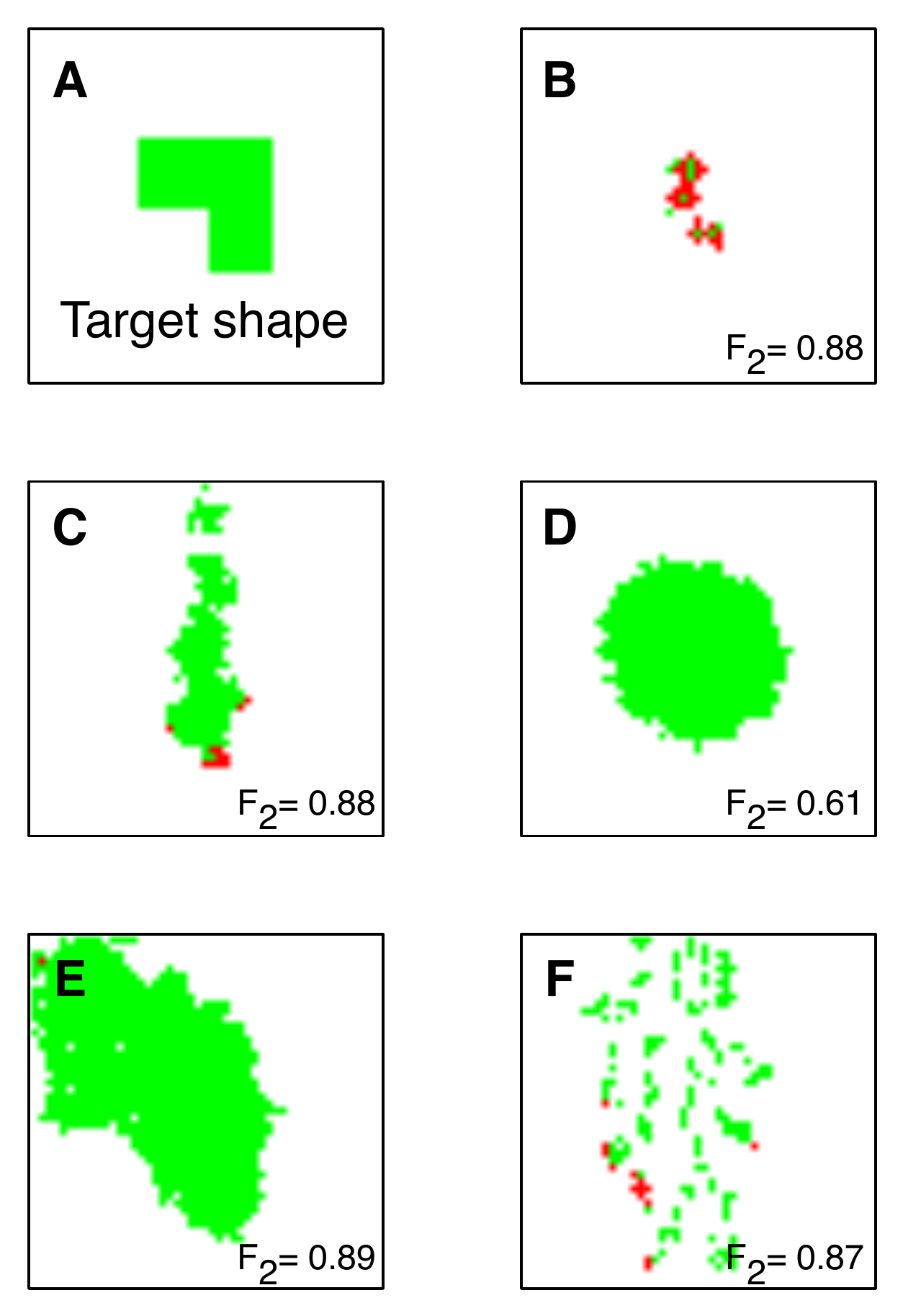}}
\caption{\label{fig:pert}Examples of target shape in a perturbed environment and their corresponding fitness $F_2$.  {All simulations are shown after 200 cell cycles have elapsed}. The binary numbers in the top right corners show the model setting (see section \ref{sec:methods}). Movies of all panels can be found at: \href{http://www.youtube.com/playlist?list=PLfj1iGvMEpeV5W2PDRCU-rIr9Q\_\_xnrGR}{http://www.youtube.com/playlist?list=PLfj1iGvMEpeV5W2PDRCU-rIr9Q\_\_xnrGR}}
\end{figure}

\noindent \textbf{Target shape.} Just as experienced by de Garis the L-shape that we were evolving for is difficult to achieve. Figure \ref{fig:unpert}B-E shows a couple of solutions to the problem. They have all entered a stationary phase (containing only quiescent cells), except panel F, which exhibits proliferating cells at the tips. While this solution makes use of cell polarisation to grow in specific directions, the other solutions avoid using it, which suggests that the generated shape could have different orientations in different realisations of the process. {The low similarity between the target shape and evolved solutions (see fig. \ref{fig:unpert} and \ref{fig:pert}) suggests that our system is not correctly formulated to evolve arbitrary shapes. This could be due both to the genetic representation and the  choice of cell dynamics.}

\subsubsection{Statistical analysis}
 Our general aim is to understand, quantitatively, how different cell capabilities contribute the ability to evolve certain target functions. The addition of a cellular trait, such as the ability to move, might make the target easier to fulfill, but at the same time it increases the space of possible solutions. In general the addition of a single output or input node increases the dimensionality of the solution space by $n=25$ parameters, since the added node is connected to $n$ other nodes whose weights need to be specified. The net benefit of adding a node is therefore the sum of its positive effects on achieving the target (LGF is for example useful when maintaining shape homeostasis) minus the cost of increasing the solution space. 

In the spirit of this cost-benefit tradeoff, we view the fitness of each solution as a linear combination of the cellular traits and make the following regression ansatz,
\begin{equation}\label{eq:regression}
F_i = \beta_0 + \mathbf{S_i}^{\rm T}\mathbf{\beta} + \varepsilon
\end{equation}
which states that the mean fitness of solution $i$ (averaged over 10 realisations) is the sum of baseline fitness $\beta_0$ plus a linear combination of the different traits, where $S_{ij} = 1$ if trait $j$ is present in solution $i$ and 0 otherwise, and $\varepsilon = (\varepsilon_1,\varepsilon_2,....)$ is an error term. $\beta = (\beta_{\rm SGF},\beta_{\rm LGF},\beta_{\rm D},\beta_{\rm M},\beta_{\rm P})$ are the regression parameters or coefficients that describe the net impact of the presence of the traits on the fitness. Before the regression was carried out we removed two types of outliers: those that were deemed unfit ($F=0$) and those that achieved perfect fitness ($F=1$). Please note that we also tried more sophisticated statistical models that included interaction terms and quadratic terms (not shown), but that the linear model \eqref{eq:regression} gave the best fit. We use the Matlab command \texttt{regress} to solve for $\beta$ and the results are presented in table 2.

\noindent {\bf Shape homeostasis.} For this target function there was 431 unfit solutions and 396 perfect solutions (out of 1200). The mean fitness across all viable solutions was $\bar{F} = 0.87$.

All the perfect solutions had $S_{\rm LGF} = 1$, and among the unfit those without LGF were significantly overrepresented (Pearson's $\chi^2$-test $p = 6 \times 10^{-10}$). This suggest that LGF has a strong positive impact on achieving shape homeostasis. From the regression parameters we can conclude that SGF, LGF and polarisation all have a positive effect on achieving shape homeostasis, and that LGF seems to be the most important. The setting with the largest mean fitness was the one with LGF and SGF and no other cellular traits. It achieved a mean fitness of $\bar{F} = 0.99954$. 

\noindent {\bf Target shape.} In this case there were no perfect solutions, which reflects the difficulty of the task. {We found 54 unfit (i.e. contained no living cells) solutions (out of 1200) and these were evenly distributed across the settings.}
The mean fitness among all viable solutions was $\bar{F} = 0.66$. From looking at the regression parameters we observe that SGF, LGF, movement and polarisation all had a positive impact on fitness. Again LGF is topping the list, which is not surprising given its ability to serve as a form of remote sensing with which the cells can gauge the total cell number in the growing structure. The setting with the highest mean fitness was 11011, i.e. the one lacking only cell death, which achieved $\bar{F} = 0.87$. 



\begin{center}
\begin{table}
    \begin{tabular}{ | l | l | l | l |} 
    \hline
    Parameter & Shape homeostasis & L-shape\\ \hline \hline
    $\beta_0$ & 0.75 & 0.39\\ \hline
    $\beta_{SGF}$ & 0.04 & 0.01 \\ \hline
    $\beta_{LGF}$ & 0.15 & 0.42 \\ \hline
    $\beta_{D}$ & -0.07 & -0.08 \\ \hline
    $\beta_{M}$ & -0.03 & 0.02 \\ \hline
    $\beta_{Pol}$ & 0.10 & 0.02\\ \hline
   
        \end{tabular} 
        \caption{ \label{tab:regress}The regression parameters of the linear fitness model \eqref{eq:regression} for the unperturbed case, obtained via least squares minimisation.}
        \end{table}
\end{center}

\subsection{Perturbed case}

We now turn to the situation in which the structures are affected by detrimental death particles that move randomly in the domain and kill cells upon contact. This changes the conditions for achieving the different target functions and as before we first analyse the results qualitatively and then move on to statistical analysis. 

\subsubsection{Qualitative analysis}

\begin{figure}[!htbp]
   \centerline{\includegraphics[width=10cm]{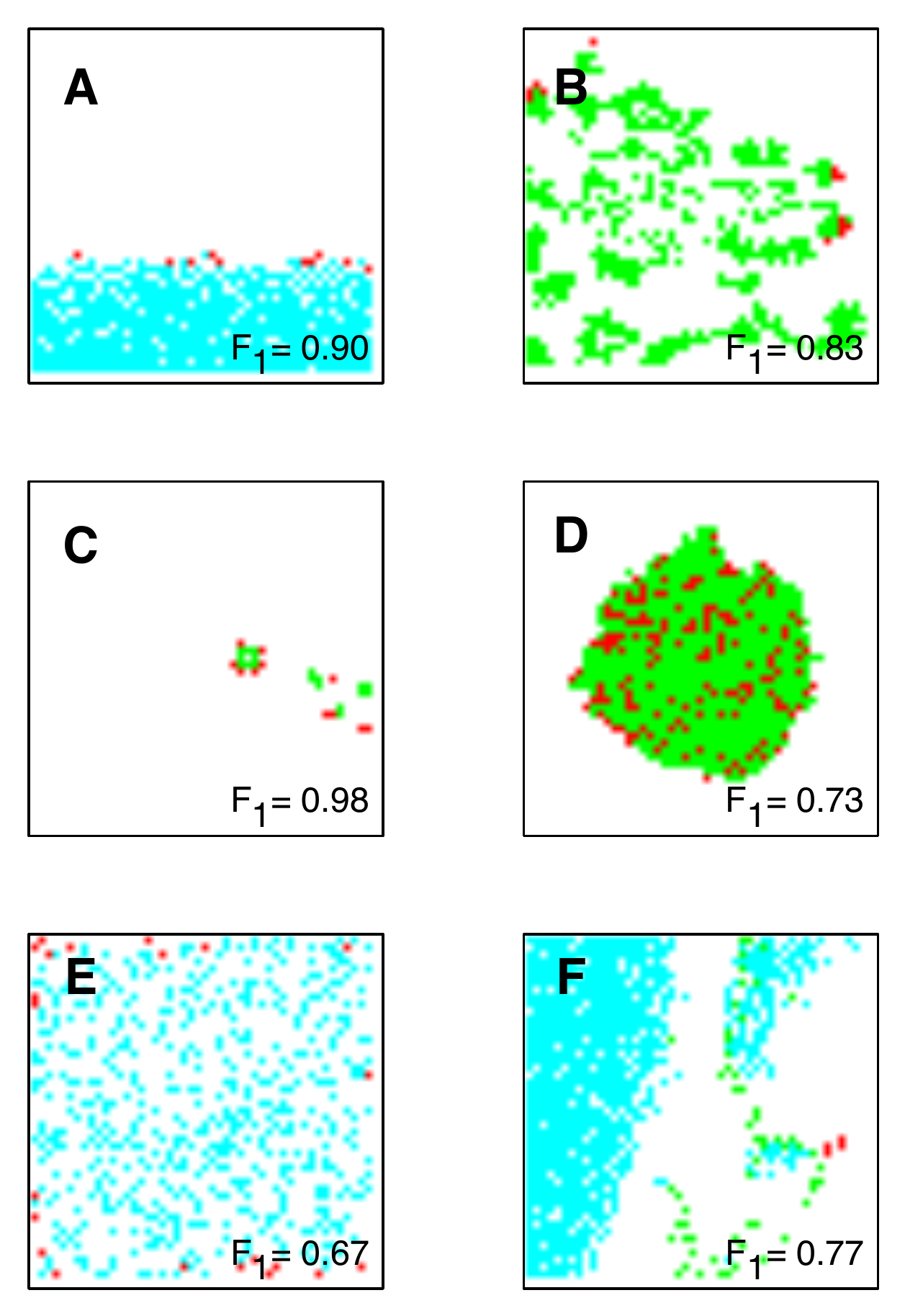}}
\caption{\label{fig:perh}Examples of shape homeostasis in a perturbed environment and their corresponding fitness $F_1$. Under these conditions all solutions need to be dynamic and replace cells killed by the harmful death particles. Despite this we see a wide variety of solutions, where some exhibit a compact morphology while other span the entire domain.  {All simulations are shown after 200 cell cycles have elapsed}. The binary numbers in the top right corners show the model setting (see section \ref{sec:methods}). Movies of all panels can be found at: \href{http://www.youtube.com/playlist?list=PLfj1iGvMEpeUdJ0bJKChBrhVFav\_YjuLt}{http://www.youtube.com/playlist?list=PLfj1iGvMEpeUdJ0bJKChBrhVFav\_YjuLt}}
\end{figure}

{\bf Shape homeostasis.} In the absence of death particles we observed two classes of solutions: static and dynamic. But with the introduction of external perturbations the former class no longer presents a possible solution since a static structure would slowly erode given the constant environmental insult. However, we still observe a number of subclasses of the dynamic solution strategy (see fig. \ref{fig:perh}). Panel A shows  a solution which forms a layer of cells at the bottom of the domain, in which most cells are migratory and only a small fraction are dividing. The lack of rotational symmetry suggests that this morphology is driven by cell polarisation, and the controlled proliferation suggests that either SGF or LGF is limiting growth. Panels B and C represent another class of solutions that exhibit a fragmented morphology. By analysing the dynamics of these solutions we find that they are driven by LGF that acts to inhibit cell division. As an island grows there is a build-up of LGF, and when it reaches a critical size cell division is inhibited and the cells turn quiescent. The death particles then start eroding the structure, but when the size falls below a certain threshold the cells again become proliferative and expand the island. Hence these structures exhibit an oscillatory behaviour. The structure shown in panel D resembles the previously discussed faux homeostasis class, which, in the presence of environmental perturbations, actually works even better since the slow expansion is almost perfectly balanced by the environmental insult. A completely different solution strategy is shown in panel E, where a sparse cell pattern fills the entire domain. Most of the cells are in a motile state, while only a few cells actively divide. This type of sparse morphology is commonly seen with quiescent cells dominating the population, however, if cell death is present then only proliferating (and dying) cells are observed. The final example (F) represents a diverse class of solution that have a non-trivial and dynamic morphology, where all cell actions: proliferation, quiescence and migration appear. In this particular example cells are generated by a travelling wave type behaviour on the right hand side of the domain. The cells that proliferate are polarised and leave their offspring in the south direction. Some of these cells turn migratory, and migrate in the opposite direction. When they reach the north boundary of the domain they move west and replenish the layer of cells on the western boundary. 

\noindent {\bf Target shape.} Again we observe that the target L-shape is difficult to achieve. It is possible to distinguish three general classes of solutions: symmetric solutions (panel D), solutions with 180 degree symmetry (B, C and F) and asymmetric solutions (E). Since the target shape lacks symmetry, the asymmetric solutions can  be said to lie closest to the desired shape, but solutions of all classes are amongst those with highest fitness. In general the solutions with 180 degree symmetry are fragmented and arise in the same way as solution B and C in fig.\ \ref{fig:perh}, with the difference that the cells are polarised either in the north-south or east-west direction.  It also worth distinguishing between dynamic solutions that contain proliferating cells and maintain their size, and those that shrinking due to the environmental insult. However, since quiescent cells can become proliferative upon changes in the multi-cellular structure, distinguishing between static and dynamic solutions requires simulations to last until all cells have been eliminated. 


\subsubsection{Statistical analysis}

{\bf Shape homeostasis.} Among the 1200 solutions we found 465 unfit solutions and as expected no perfect solutions. The mean fitness was $\bar{F} = 0.84$, which was significantly smaller than the unperturbed case (two-sample t-test, $p < 0.002$). The regression parameters, shown in table 3, reveal that the impact of the cellular traits is similar to the unperturbed case, the biggest difference being the impact of SGF, which in the perturbed case contributes 40 times less to the fitness. A similar trend is seen with polarisation, but in this case the difference is only three fold. Again LGF is the most important characteristic, and the highest expected fitness is achieved under setting 11001, which suggests that the addition of cell death and migration not only increases the size of the search space, but adds little towards achieving shape homeostasis. 

\noindent {\bf Target shape.}  We found more than twice as many unfit solutions when perturbations are present (132 vs.\ 52), but the mean fitness of the viable solutions ($\bar{F} = 0.66$) was almost identical to the unperturbed case. This is somewhat surprising since the development in a harmful environment would be expected to be a more difficult task. {However, if we look at the regression parameters (table 3) we see that all characteristics except LGF now have a negligible impact on fitness.}
One would expect that the most fit setting is 01000, but in fact 11000 has a slightly higher mean fitness (0.76 vs. 0.81). The presence/absence of LGF is a strong determinant of the fitness (the mean fitness for LGF on is 0.76, while it is 0.49 for LGF off).


\begin{center}
\begin{table}\label{tab:regress}
    \begin{tabular}{ | l | l | l | l |} 
    \hline
    Parameter & Shape homeostasis & L-shape\\ \hline \hline
    $\beta_0$ & 0.73 & 0.57\\ \hline
    $\beta_{SGF}$ & 0.001 & -0.03 \\ \hline
    $\beta_{LGF}$ & 0.16 & 0.25  \\ \hline
    $\beta_{D}$ & -0.04 & -0.07  \\ \hline
    $\beta_{M}$ & -0.04 & 0.005 \\ \hline
    $\beta_{Pol}$ & 0.03 & -0.04\\ \hline
   
        \end{tabular} \caption{ \label{tab:pert}The parameters of the linear fitness model \eqref{eq:regression} for the perturbed case, obtained via least squares regression.}
        \end{table}
\end{center}

\subsection{Robustness analysis}
All developing systems are subject to more or less noise, both in terms of internal fluctuations such as mutations and external changes such as toxic agents that disrupt growth. The ability of real systems to withstand such perturbations suggest a large degree of robustness and we want to investigate if this is naturally present in our evolved solutions as well. 

{Previously we considered the effect of external perturbations in the form of harmful particles that kill the cells, but what if the magnitude of the perturbations is much larger or if it is not external, but internal to the cell? In order to answer these questions, and assess  the robustness of the evolved solutions we subject them to extensive wounding (large external perturbation) and/or mutations (internal perturbations).} The former scenario is achieved by killing each cell in the simulation after 100 time steps, with probability $p_w=0.75$ (i.e. on average 75\% of the cells are eliminated), and the latter is implemented by allowing for mutations to occur during cell division with probability $p_m = 0.01$ per matrix element. {If a matrix element is chosen to be mutated a random number drawn from a normal distribution with mean 0 and variance 0.7 is added to the existing matrix element.}

We quantify the effect of these perturbations by measuring the viability of the solutions in the presence of wounding, mutations and both perturbations acting simultaneously. The viability $W$ we define here as the fraction of simulations in which a solution achieves non-zero fitness. This measure also accounts for the fact that the perturbations are random by nature and therefore their effect needs to be averaged. The $W$'s are calculated by simulating the effect of the perturbations across 50 independent simulations.

In order to analyse the combined effect of the perturbations we make use of the Bliss interaction model, which is used for assessing drug-drug interactions, and gene epistasis \cite{lee2007}. For a given solution one calculates the interaction score
\[
\varepsilon = W_{wm} - W_{w}W_{m}
\]
where $\varepsilon = 0$ indicates no interaction (i.e.\ $W_{wm} = W_{w}W_{m}$), $\varepsilon < 0$ means that the combined perturbation has a larger effect than expected ($W_{wm} < W_{w}W_{m}$, usually termed synergism), and $\varepsilon > 0$ suggest a buffering effect termed antagonism. 

\begin{center}
\begin{table}\label{tab:epi}
    \begin{tabular}{ | l | l | l | l | l |} 
    \hline
    Setup & $\bar{W}_w$ & $\bar{W}_m$ & $\bar{W}_{wm}$ & $\bar{\varepsilon} = \bar{W}_{wm}-\bar{W}_w \bar{W}_m$ \\ \hline \hline
    Stable--any shape & 0.96 & 0.94 & 0.91 & 0.0076\\ \hline
    Stable--L-shape & 0.98 & 0.99 & 0.98 & 0.0098\\ \hline
    Unstable--any shape & 0.94 & 0.89 & 0.86 & 0.0043\\ \hline
    Unstable--L-shape & 0.97 & 0.99 & 0.98 & 0.0175\\ \hline
   
        \end{tabular} \caption{\label{tab:epi} The mean viability for the different setups in the case of wounding ($\bar{w}_w$), mutations ($\bar{w}_m$) and both ($\bar{w}_{wm}$).} 
        \end{table}
\end{center}

In this section we restrict our perturbation analysis to solutions that are, in the absence of wounding and mutations, perfectly viable (i.e.\ all have $W=1$). Table \ref{tab:epi} summarises the effect of the perturbations on viability and also their combined effect. In general we see that the perturbations have a small effect on viability which lies above 90\% in all cases. It is also interesting to note that the solutions that had evolved in the unstable environment did not do better when it came to wounding. 

\begin{figure}[!htb]
   \centerline{\includegraphics[width=9cm]{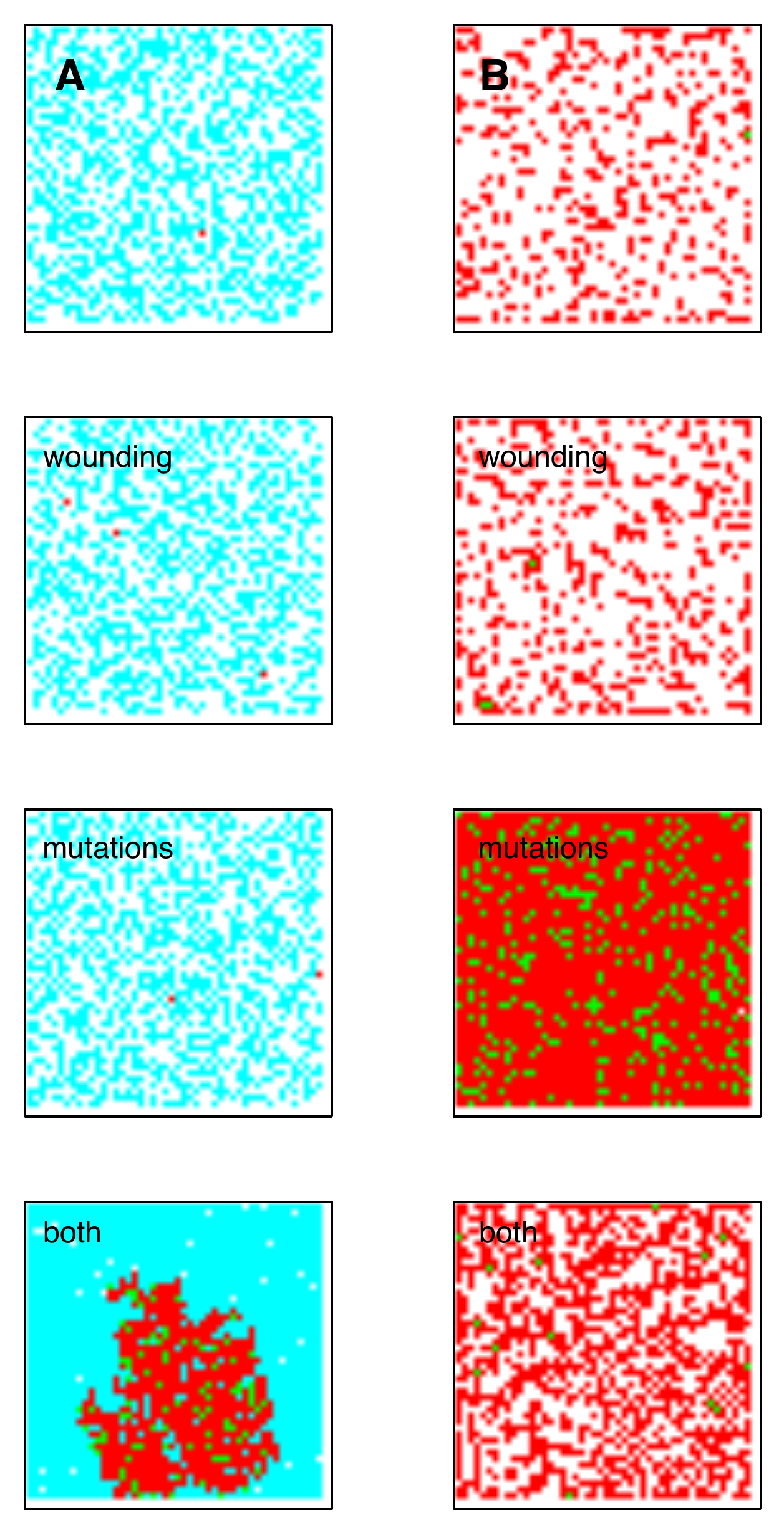}}
\caption{\label{fig:perturb}Examples of two solutions in which the perturbations act in a synergistic (A) and antagonistic way (B).  {All simulations are shown after 200 cell cycles have elapsed}. Movies of all panels can be found at: \href{http://www.youtube.com/playlist?list=PLfj1iGvMEpeUBBsiJEZ-s6f\_Dn6baK2MQ}{http://www.youtube.com/playlist?list=PLfj1iGvMEpeUBBsiJEZ-s6f\_Dn6baK2MQ}}
\end{figure}

Figure \ref{fig:perturb} shows two examples of a synergistic interaction (A) and antagonistic one (B). In the former case the combined effect on viability of wounding and mutations ($W_{wm}=0.36$) is stronger than expected from the interaction model ($W_m \times W_w = 0.42$). A possible explanation is that the wounding increases the number of cell divisions, which increases the likelihood of a detrimental mutation that could disrupt the homeostatic process. The antagonistic case (fig.\ \ref{fig:perturb}b) is more difficult to explain, but also in this case it is mutations that have the most disruptive effect. Here, however, the combined effect is smaller ($W_{wm}=0.76$) than mutations on their own ($W_{m}=0.56$). 

In general it was difficult to discern any obvious pattern in the response to these perturbations, however we did find that synergistic solutions ($\varepsilon < -0.1$) in the stable environment were more likely to have active LGF ($\chi^2$-test $p < 0.001$ and $p < 4 \times 10^{-4}$ for any shape and L-shape respectively). 

Further we looked at connections between the morphology of the structure (size and cell turn-over rate) and the response to perturbations.
In the case of shape homeostasis in the stable environment there was a negative correlation between viability under mutations and cell turn-over rate ($\rho = -0.70$). This observation did not persist in the other cases, but solutions to shape homeostasis in the unstable environment that had zero viability under mutations had a higher cell turn-over (ranksum test, $p<0.05$). 
Lastly we expected smaller structures to be more sensitive to wounding, but the data did not support this conclusion.

\subsection{Trading places}
{As a final test of robustness we evaluated the viability of the evolved solution in the opposite environment. This should give us some idea of how plastic or adaptable the solutions are to new growth conditions. The fitness of solutions evolved in the stable environment was evaluated in the presence of death particles and the solutions evolved in the unstable environment were evaluated in the absence of such particles. We quantify this by calculating the average fitness difference $\Delta F = F_{t}-F$, where $F$ is the fitness in the native environment and
$F_t$ is the fitness in the opposite environment.  We found that $\Delta F = -0.36$ and $-0.25$ for the solutions evolved in a stable environment for the shape homeostasis and L-shape respectively. For the solutions evolved in the unstable environment the corresponding change in fitness was -0.07 and 0.07. This suggests that the latter solutions were better at adapting to growth conditions they were not evolved in.}

\section{Discussion}
In this study we used an individual-based model, that incorporates key cellular characteristics, defined by evolving intra-cellular networks (that control cell behaviour), to develop multicellular structures that exhibit a large range of behaviours. Focusing on the scale of multicellular structures (and currently disregarding the intracellular dynamics) we observe chaotic patterns (driven by the randomness in the placement of daughter cell, e.g.\ fig.\ \ref{fig:unperh}C), patterns growing linearly in time (fig.\ \ref{fig:unperh}E), patterns with non-trivial steady states (fig.\ \ref{fig:unpert}C), diffusion limited growth (fig.\ \ref{fig:unperh}B) and oscillatory dynamics (fig.\ \ref{fig:perh}F). Explaining the development and dynamics of these behaviours would require a substantial amount of work that is beyond the scope of this paper. Essentially it would call for a detailed analysis of the intra-cellular signalling networks, coupled to the cell-cell communication facilitated by SGF and LGF. The behaviour of each class may then be reduced to a simpler model and better explained. For example the tesselation produced in fig.\ \ref{fig:unperh}A suggests an interesting interplay between short- and long-range inhibition. 
Such a strategy of simplification and reduction was successfully applied to \textit{Dictyostelium discoideum} \cite{sgro2014}, where the intra-cellular dynamics were modelled using the FitzHugh-Nagumo model. Their approach was motivated by previous work in the physics of critical phenomena \cite{kadanoff2000}, where a small number of \textit{universality classes} have been identified.


{We were unable to evolve a single solution (out of 1200) that fully recapitulated the target shape. This could be due to our choice of genetic representation or choice of cell actions.} However,  evolving an L-shape presents a special difficulty since it requires the symmetry of the growing structure to be broken twice. A rectangle (with uneven sides) would require a single symmetry breaking, but still exhibits 180 degree symmetry, whereas the L-shape has no rotational symmetry. de Garis \cite{DeGaris1992} who tried to evolve precisely this shape, solved the problem by allowing for two distinct rule sets (operons) to be used by the cells. One operon would be used for growing the horizontal arm of the L and the other for growing the vertical arm. The two rule sets and the rule for switching between the two were then found using a genetic algorithm. This option is not explicitly available in our modelling framework, although it is still possible for cells to exist in two distinct differentiation states, such that they respond differently to identical stimuli. Evolving such a behaviour turned out to be difficult and most solutions only managed to break symmetry once and obtain 180 degree symmetry (e.g.\ fig.\ \ref{fig:unpert}C). Although, the structure shown in fig.\ \ref{fig:unpert}B did in fact achieve double symmetry breaking, but upon repeated simulations it was noted that the shape produced is not preserved, and that random chance (the division of unpolarised cells) influences the final shape. Another interesting solution is shown in fig.\ \ref{fig:unpert}F, which exhibits a perfect four-fold (90 degree) symmetry.  Here the dividing cells reside on the tips of the protruding branches, much like the stem cells in plants which are located in the meristem \cite{Weigel2002}. {It could be that the poor outcome is due to our fitness function being to rigid. If we allow for translation and/or rotation of the target shape we might achieve a higher average fitness. This is in fact sensible from a biological point of view since the fitness of a certain structure might be insensitive to its exact position and orientation.}

The range of behaviours observed in the current model highlights the problem of understanding the connection between the properties of a model (or model setting in our vocabulary) and the patterns it can give rise to. For example it is obvious that the cross-like pattern shown in fig.\ \ref{fig:unpert}F requires (or rather is extremely unlikely in the absence of) polarisation, but for other patterns the answer is not obvious.  
Given the algorithmic structure of the model it might be possible to formalise these questions and mathematically prove what type of structures are attainable within a certain model. Some inroads into a similar problem, but in the context of collective construction, have recently been made by Werfel and co-workers \cite{Werfel2014}. Given a certain structure they were able to prove if the structure was attainable in their system, and also reverse-engineer the microscopic rules required to build it.

The most obvious conclusion from this study is that LGF (long range diffusing growth factor) has a large impact on the outcome of the evolutionary algorithm. When the cells are able to produce and sense this diffusible factor (which can influence any other trait, not just cell division) the expected fitness is significantly higher for both the shape homeostasis and the target L-shape. This is unsurprising since LGF provides a means of the cells to perform quorum sensing (i.e.\ ``count'' the total number of cells in the multi-cellular structure). The effect of using LGF is however quite varied and it can induce fragmented, branched and circular morphologies (see fig.\ \ref{fig:unperh}a, b and f), depending on the architecture of the intra-cellular network. {It is worth noting that the structures that use LGF rely on the no-flux boundary conditions which is the reason why there is a build up of LGF in the domain. The reliance on the boundary conditions is in fact a general feature of many evolved solutions, which use the bounded domain as a means to stop growing. However, homeostasis induced by the bounded domain is not possible in our model since we assign zero fitness to all solutions that fill the entire domain with cells. }

The difference between the cellular structures evolved in the unperturbed and perturbed environments is most evident in the case of shape homeostasis. In the latter case, 80 \% of the solutions were static:  they developed into a fixed structure containing only quiescent cells. When evolving in a harmful environment this strategy is no longer a possible solution, and shape homeostasis must be achieved in a dynamical fashion. Intuitively one would assume that this harder to achieve, and indeed we observe consistently lower fitness in the perturbed case across all 48 settings. In addition, the mean fitness within each setting in the unperturbed and perturbed case are strongly correlated ($\rho = 0.92$), suggesting that the effect is uniform across all settings. In the target shape case we also observe a strong correlation between the mean fitness in the different classes in the perturbed and unperturbed environments ($\rho = 0.91$), but here the presence of LGF tips the scale. When LGF is turned on, the unperturbed solutions do better, while in its absence the perturbed solutions have a higher mean fitness. 

We have also investigated the robustness of the evolved solutions with respect to wounding and mutations. In general the solutions were  robust to perturbations with a mean viability of above 90 \%, and surprisingly, there was no significant difference between the stable and unstable environment. By analysing the data using the Bliss model we could also show that for most solutions the combined perturbations acted as if they were independent. Among the solutions for which the combined perturbation had a larger effect than expected (i.e. synergistic effect) settings involving LGF were overrepresented.
In the case of shape homeostasis we could also show a negative correlation between a high rate of cell turnover and viability in the presence of mutations. This is to be expected since each cell division increases the risk of mutations that could give rise to a mutant which disrupts the homeostatic structure. This is similar to tumours that most often appear in epithelial tissues which have a high rate of cell turnover \cite{Noble2015}. 

Despite the similarity between the solutions from the stable and unstable environments with respect to perturbations, there was an obvious difference when the solutions were tested in environments for which they were not selected for. Solutions from the stable environment did much worse in the unstable setting than did solutions from the unstable in the stable environment. In fact 25\% of the solutions to shape homeostasis in the unstable environment achieved perfect homeostasis in the stable environment. These solutions are thus truly repairing the damage induced by harmful death particles, and do not grow at a rate which is balanced by the environmental insult. This suggests a complex regulatory mechanism where cell death is sensed by surrounding cells that repair the damage.

The current study focuses on the level of collective cellular behaviour, and our aim has been to understand which cellular traits facilitate the formation of certain morphologies. However, in order to fully comprehend the dynamics of the multicellular structures we have evolved, one would first have to analyse the behaviour of the individual intracellular signalling networks, and then move onto analysing how cell communication (via SGF, LGF) and cell actions (migration, death and proliferation) interact and give rise to multicellular dynamics. This is a challenging task and not within the scope of the work we present in this paper. Given existing techniques for analysing intracellular networks (genetic regulation) and cell-cell communication it would be more realistic to dedicate a separate paper to examining the behaviour of the travelling wave system (fig.\ \ref{fig:perh}F) for example. 

We have created a plethora of basic multi-cellular life forms, 
and understanding them presents similar challenges to those posed by real biology. In fact the evolved structures could be used as a testing ground for techniques and metrics that may be useful in increasing our understanding of biology. A possible approach to this problem is to identify minimal models (i.e. networks and traits) characterising  each class of behaviour (i.e. universality classes). This would yield a catalog of models that could then be used for interrogating real systems that exhibit similar behaviour.  Even though we are far from this goal, the results presented so far provide some insights into what traits are important for multicellular organisation. 

Our results suggest that adding traits has both a positive and negative effect on the ability to form complex morphologies, and we have conjectured that the negative effects are related to the increase in the size of the search space. The genetic algorithm is only run for 20 generations with a population size of 25 candidate solutions, which means that only a fraction of the genotype space is ever explored. When a new trait is added the size of the genotype space typically increases linearly with the number of nodes in the intra-cellular network (the new node is connected to all existing nodes), which makes our EA even less efficient. A means to test this hypothesis would be to make use of some sort of exhaustive search algorithm. If negative effects still persist in that scenario the trait in question has a truly negative impact on the ability to form certain morphologies. In relation to real evolving systems it is worth noting that the genotype space of real organisms does not remain constant in size, but is constantly changing under the forces of selection. Implementing this feature in our framework could potentially make it much more efficient. {A complementary approach would be to simulate incremental evolution, where traits are added incrementally during the course of the evolutionary algorithm. In a biological setting this would correspond to increasing the size of the genetic network, and hence to an increase in the size of the genome. An interesting question in this context would be what sequence of traits is most likely to yield a homeostatis structure. Alternatively one could add and remove traits with a small probability and let the evolutionary algorithm find the most likely sequence of traits.}

{The conclusions drawn from this study are naturally dependent on our modelling choices. One could imagine variations of our framework in which the cells are not constrained to a lattice and instead interact mechanically, or simulations performed on an unbounded domain where the boundaries cannot be used to facilitate a homeostatic structure (e.g. the build-up of LGF in the domain). It is also possible to consider other models of the intra-cellular dynamics, such as boolean networks \cite{Kauffman1969}, that are known to exhibit distinct attractors that correspond to different differentiation states  in the cell. One could also consider continuous time version where the state of the gene network is determined by a set of coupled ordinary differential equations. The degree to which these choice affect the dynamics and hence conclusions of our work is something we plan to investigate in the future.}

{ The number of fitness targets (arbitrary homeostasis and the L-shape) is quite limited and more could probably be learnt about the impact of cellular capabilities if a wider range of fitness targets were considered (e.g.\ limb-like patterns or patterns of varying degrees of symmetry). For example it is known that cell death and polarity plays a crucial role in homeostasis and repair in flatworms \cite{Pellettieri2010}. In our analysis these two traits were given little importance. Is this due to simplistic fitness targets or other modelling choices? We hope to return to these questions in future work.}

\section{Conclusion}
In this study we have focused on the impact of cellular traits on the evolution of collective behaviour in a multi-cellular context. This work was carried out with the help of an agent-based model that covers many previous models in the field of artificial life (see table \ref{tab:sum}). This allowed us to asses the subset of capabilities required for efficient morphogenesis and homeostasis. {Our results suggest that in particular long range cell-cell communication positively influences the ability to achieve shape homeostasis and growing into an L-shape, but we also observe that short range communication and cell polarisation have a positive effect.} 
We have also shown that solutions that evolve in an unstable environment are more dynamic and robust with respect to changes in external conditions, and thus conclude that a model with SGF, LGF and cell polarisation is better positioned to produce a successful solution. {Our approach to an unstable environment was to include harmful particles, but one could also consider other means of perturbation such as noise in the signalling networks or perturbations to the growth factors dynamics.} We hope that this first systematic study of collective behaviour in morphogenesis will inspire future work that will be able to better classify and analyse the vast range of dynamic behaviours we have reported in this paper.

\end{spacing}
\end{document}